\documentclass[10pt]{wlscirep}
\title{Mechanism of pressure sensitive adhesion in nematic elastomers}

\author[*]{Hongye Guo, Mohand O. Saed  and Eugene M. Terentjev}
\affil{Cavendish Laboratory, University of Cambridge, Cambridge, CB3 0HE, U.K.}
\affil[*]{emt1000@cam.ac.uk}

\begin{abstract}
Nematic liquid crystal elastomers (LCEs) have anomalously high vibration damping, and it has been assumed this is the cause of their anomalously high pressure-sensitive adhesion (PSA). Here we investigate the mechanism behind this enhanced PSA by first preparing thin adhesive tapes with LCE of varying crosslinking density, characterizing their material and surface properties, and then studying the adhesion characteristics with a standard set of 90-deg peel, lap shear, and probe tack tests. The study confirms that the enhanced PSA is only present in (and due to) the nematic phase of the elastomer, and the strength of bonding takes over 24 hours to fully reach its maximum value. Such a long saturation time is caused by the slow relaxation of local stress and director orientation in nematic domains after pressing against the surface. We confirm this mechanism by showing that a freshly pressed and annealed tape reaches the same maximum bonding strength on cooling, when the returning nematic order is forming in its optimal configuration in the pressed film. 
\end{abstract}
\begin{document}

\flushbottom
\maketitle
\thispagestyle{empty}

\section{Introduction}

Liquid crystalline elastomers (LCEs) are a class of functional materials that attract attention in many different applied fields. Originally, the main interest was in utilizing LCEs as soft actuators with a very high stroke, and actuation stress capable of reaching the MPa scale.\cite{Kupfer1991,Tajbakhsh2001} The discovery of `soft elasticity' in nematic LCEs has led to another strand of potential applications utilizing their anomalously high viscoelastic dissipation and vibration damping \cite{2Tajbakhsh2001,Merkel2019}, although it has recently been suggested that the enhanced damping is not actually related to the macroscopic soft elasticity but is controlled by the high-frequency rotation of nematic director on the microscopic, sub-domain scale \cite{SciRep2023}. Recently, there have been several reports demonstrating a new effect of an enhanced pressure sensitive adhesion (PSA) on the surface of nematic LCEs \cite{Ohzono2019,Hiro2022}, alleging a direct link between the viscoelastic dissipation in the material, measured by the loss factor $\tan \delta (\omega)$, and the effective energy of surface adhesion $\gamma$. That is, similar to many other polymer-based PSA systems \cite{Creton2006,Creton2009,Israelachvili2016,Creton2020,Shull2020}, the strength of adhesion is dominated by a physical factor additional to the `ordinary' surface energy; this factor is dominated by the integral of  $\tan \delta (\omega)$ over the range of frequencies determined by the pull-off speed \cite{Persson2004,Persson2005}. As the mechanisms of PSA are generally not very well understood, even in spite of several good theoretical approaches \cite{DeGennes1996,Raphael2004,Doi2018}, and since in LCEs we have an additional strong effect of nematic director relaxation, here we investigate the enhanced LCE adhesion in some detail in order to ascertain its mechanism. 

There are several factors one has to consider in the PSA process, and in considering them we wish to exclude the chemically-defined surface tension $\gamma_0$ from this list:  we will work with a particular fixed chemistry of an elastomer, so that $\gamma_0$ remains constant (which we measure from the contact angle analysis of water droplets to be $\gamma_0 = 42$\,mN/m). The methods of estimation of surface energy of polymers have evolved since Owens and Wendth \cite{Owens1969}, but they all invariably give the range of $\gamma_0$ between 20 and 60\,mN/m (with water having the high $\gamma_0=$72\,mN/m). This is about 4 orders of magnitude lower than the measured adhesion energy $W$   {(often called the `work of adhesion')} of `sticky surfaces' in many cited studies, which is in the range of 100's\,N/m, to 1000's\,N/m \cite{Creton2020}.  The first key factor is the roughness of the polymer surface (assuming for simplicity that the pressing surface is rigid and smooth), which has been extensively studied over the years \cite{Greenwood1966,Greenwood2008}. As we will be working with a natural polydomain nematic LCE, it turns out that its free surface has an unavoidable roughness on the micron scale, reflecting the equilibrium nematic domain structure that causes the corresponding local deformation pattern \cite{Clarke1998a,Fridrikh1999,Biggins2012}.   {We just mentioned the assumption that the pressing surface is smooth, which is of course incorrect, and the famous Dahlquist criterion has been developed a long time ago \cite{Dahlquist} to relate the natural roughness of that surface to the adhesion strength: it is based on how well the PSA polymer penetrates the rigid rough surface profile, and shows that the typical rubber modulus needs to be below 0.1 MPa for a typical solid surface with single-micron roghness topography. Here is another common confusion, in assuming that better penetration increases the surface contact area which is the cause of better adhesion -- when an easy estimate shows that the change in effective contact area could be less than 5\%. It is the `work of adhesion' \cite{Creton2006,Creton2020} that is behind the empirical PSA effect.}

 The second important factor that affects PSA is the soft elasticity of polydomain nematic LCE.   {Here we send the reader to the extensive original literature, well summarized in the book \cite{LCEbook}, to understand what this novel effect unique to LCE is. Soft elasticity manifests in different ways in the tensile mode in the peel test and in the lap shear test, but in all cases leads to an enormous ductility of the LCE material.} Unlike in ordinary isotropic polymer PSA systems, the LCE in a loaded contact with substrate will be able to gradually adjust and re-align its nematic domains near the surface, following the local soft trajectory, to significantly lower the overall free energy of a bonded state. As a result, the thickness of the ``affected layer'' (which is about 10-100\,nm in ordinary polymer networks \cite{Abbott}) extends to several microns.   {The elastic energy stored in this layer is our proposed mechanism of the enhanced adhesion of polydomain LCE, as discussed towards the end of this paper, once we get familiar with the measured properties of this experimental system.}

The two factors discussed above are related to the third characteristic feature of polydomain LCE, having a profound impact on PSA properties. The relaxation dynamics of polydomain nematic LCE is extremely slow \cite{Clarke1998b}. Therefore, we expect that the contact time during PSA bonding would have a very strong effect on the strength of adhesion in LCE, allowing the sufficient domain re-organization and adjustment under pressure against the substrate.   {This would be highly unusual for the ordinary isotropic-pgase PSA systems.} And finally, nematic LCEs are renowned for their anomalous viscoelastic dissipation, measured by the high loss factor  $\tan \delta (\omega)$  \cite{2Tajbakhsh2001,Mistry2021,ncom2021}, which arises not because of the `usual' lowering of the storage modulus (as is the case in lower crosslinking densities or melts), but due to the additional mechanism of internal dissipation due to local rotational motions of nematic director, controlled by its rotational viscosity, which is in turn controlled by the thermally-activated exponential factor of overcoming the mean-field barrier for mesogen rotation \cite{Osipov,Chan2007}. Both disappear in the isotropic phase of LCE, returning  $\tan \delta $, and the PSA strength, to ordinary elastomer values. 

  {The first study that has mentioned the PSA in LCE is the theoretical work of Corbett and Adams \cite{Corbett2013}, much earlier than its first experimental finding \cite{Ohzono2019}. This study applied the equilibrium `Trace formula' for nematic elasticity \cite{LCEbook} and the dynamic dumb-bell rheological model of Mafettoni and Marucci \cite{Marucci} to explore how the the nematic rotational degree of freedom affects the PSA layer debonding via cavity nucleation and then fibrillation. They found that in the director geometry when the rotation would accompany local deformation, the effective work of adhesion was higher than when there was no underlying rotation (the director perpendicular to the adhesion surface in the flat probe tack test). However, they also found that the isotropic-phase adhesion was as high as the parallel-alignment adhesion, which is strongly contrary to all observations. In any case, their pioneering work has little to do with out case, when we deliberately work with the practically relevant non-aligned polydomain LCE layer, and the local inhomogeneous deformations in it are behind our proposed mechanism of enhanced adhesion.}

In this paper we  carry out several technology-standard adhesion tests to assess how the key factors discussed above affect its strength and other features. Here we do not aim to `optimize' the polymer system, i.e. find the best LCE adhesive material or compare with other non-LCE PSA,  which should be a subject of a separate study of the materials chemistry. Instead, we work with a standard well-characterized LCE material,   {described and characterized in dozens of papers,} and focus on the adhesion mechanism itself,   {which is universal and does not depend on the chemical nature of the material}. Similarly, we do not explore different adhesion surfaces, but stay consistently with the case of LCE adhesion to glass, in different configurations and settings.

 %%%%%%%%%%%%%%%%%%%%%%%%%%%%%%%%%%%%%%%%%%%%%%%%%
\section{Experimental section}

\subsection*{LCE film preparation}
The starting materials for our LCE are all commercially available and relatively cheap. To produce the `standard' reference LCE materials, we followed the established synthesis \cite{Saed2015,Saed2017} using the Michael addition click reaction between RM257 reacting mesogens (4-(3-acryloyloxypropyloxy)benzoic acid 2-methyl-1,4-phenylene ester di-acrylate) obtained from Daken Chemical Ltd. and the EDDT thiol spacers (2,2'-(ethylenedioxy) diethanethiol). Butylated hydroxytoluene inhibitor (BHT) was used to prevent acrylate homopolymerization, and Triethylamine (TEA) is one of many that work well with thiol-acrylate chemistry. We specifically chose the weak TEA catalyst (instead of a much stronger catalyst such as Dipropylamine) to have a low reaction rate, in order to obtain a narrower and consistent chain length distribution.

We controlled the crosslinking density by first preparing thiol-terminated oligomer chains of LC polymer, running the above reaction with an excess thiol. The di-acrylate mesogen RM257 was dissolved in toluene (70 wt\%) together with BHT (2 wt\%) assisted by gentle heating and vigorous mixing by a vortex mixer. Next, the di-thiol spacer EDDT was added to the same vial. After obtaining a fully dissolved, homogeneous solution, TEA (1 wt\%) was added while stirring to catalyze the reaction. The vial was then covered with aluminium foil to avoid light exposure and placed on a roller for 48 hours. There is a straightforward correspondence between the excess of functional groups during chain-extension by alternating di-functional monomers, which we have verified by running gel permeation chromatography (GPC) on these oligomers and finding a good match with the basic calculation (see \cite{Guo2023} for details of GPC analysis). In this way, we had oligomer chains of the average 20-mesogen length for the 1.05:1 thiol excess over acrylate, of the average 10-mesogen length for the 1.1:1 thiol excess, and of the average 5-mesogen length for the 1.2:1 thiol excess. The longer-chain oligomers had the appropriately much higher viscosity. The vinyl-functional crosslinker was mixed into the oligomer, together with the photo-initiator Irgacure I-369 (1 wt\%), to induce a rapid UV-stimulated crosslinking of the thin film after its spreading on the backing plastic substrate. The exact amount of the vinyl crosslinkers was determined stoichiometrically by balancing the overall thiol and acrylate groups.

We used 4-functional (2,4,6,8-Tetramethyl-2,4,6,8-tetravinyl cyclotetrasiloxane)\cite{Saed2020} and 3-functional (1,3,5-triallyl-1,3,5-triazine-2,4,6(1H,3H,5H)-trione)\cite{Saed2019} versions, and verified that the film properties after crosslinking were very similar, both mechanically and adhesively. This is because the elastomer properties are determined by the length of network strands, and only very weakly by the different crosslink functionality. For this reason, we refer to the three types of LCE material as 5\%, 10\% and 20\% crosslinked, reflecting the respective 20-, 10- and 5-mesogen long chains in the network. The weight percentages shown here are all calculated with respect to the total weight of the elastomer without solvent. All chemicals, apart from RM257, were obtained from Sigma Merck. 

The spreading of the pre-crosslinking oligomer was done using the Tape Casting Coater (MSK-AFA-III), from MTI Corporation, using the PET backing films of two thicknesses: the 23 $\mu$m film from Hi-Fi Industrial Film Ltd, and the 200 $\mu$m overhead transparencies (for the lap shear testing when the backing tape must remain undeformed under tension). The PET films were plasma-treated using the plasma cleaner from Diener Plasma GmbH \& Co. KG to ensure strong bonding on the adhesive LCE layer. After thickness-controlled spreading (500 $\mu$m), the film was exposed to UV for 1 hour for full crosslinking. The coated film was placed in a vacuum oven at 70 $^{\circ}C$ overnight for solvent removal. After preparation, the exposed surface of LCE was covered by silicone paper (we used the backing paper of Avery labels) to protect it from contamination by atmospheric factors and accidental touch. The final thickness of the homogeneous polydomain LCE layer on PET was ca. 200 $\mu$m.

\subsection*{LCE characterization}
This class of nematic LCEs has been studied extensively, and its bulk properties are well established.\cite{Saed2015,Saed2017} For instance, the glass transition $T_g$ is ca. 0$^{\circ}C$ and the nematic-isotropic transition $T_\mathrm{NI}$ is ca. 60$^{\circ}C$, although the latter could be adjusted within the range of 30-100$^{\circ}C$ by minor modifications of monomers and spacers used.\cite{Saed2017}

The equilibrium (chemical) surface energy $\gamma_0$ was measured via the contact angle, following the Owens-Wendth method \cite{Owens1969}, using the water droplet on clean glass as a reference. Our LCE is less hydrophobic than typical polyolefins, and its surface energy was found to be close to Nylon: $\gamma_0 \approx 42$\,mN/m. 

The tensile stress-strain curves were measured using the Tinius-Olsen ST1 tensiometer, with LCE made into the standard dogbone shapes (ASTM D412). Experiments were conducted at ambient temperature in the nematic phase, and also at a high temperature of ca. 90$^{\circ}C$ in the isotropic phase, for comparison. The enormous ductility of the nematic LCE is in sharp contrast with the brittle network in the isotropic phase: this finding will be key to our explaining many observations below. 

\begin{figure}
    \centering
   \includegraphics[width=\textwidth]{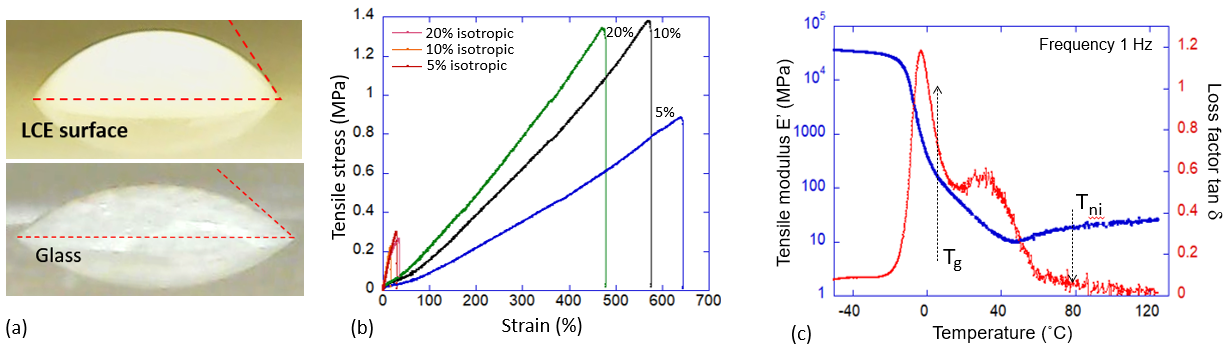}
    \caption{(a) The contact angle illustration. (b) Tensile test results for the three nematic LCE materials with different crosslinking densities; the red curves mark the tests at high temperature, in the isotropic phase. (c) DMA temperature ramp data for the 10\%-crosslinked LCE tape, at constant 1\,Hz, showing the storage modulus $G'$ and the loss factor $\tan \delta$ in three phases. $T_g$ and $T_\mathrm{NI}$ are labelled in the plot.}
   \label{fig1}
\end{figure} 

Dynamic mechanical (DMA) temperature ramp was conducted using the TA Instruments DMA850 in tension film mode. For these tests, two PET-based LCE adhesive tapes were stuck to each other face-to-face, so that the actual test of the deformable LCE was in the simple shear mode. The temperature was ramped from -40$^{\circ}C$ to 120$^{\circ}C$ with the sample oscillating at 0.1\% strain at 1Hz, and its storage and loss moduli (with tan$\delta=G''/G'$)  recorded.  This test clearly identifies the low-temperature glass, the high-temperature isotropic, and the ambient-temperature nematic LCE phases, the latter characterized by the anomalously high loss factor, see Fig. \ref{fig1} for details.

\subsection*{Probe Tack 2 test}

The Probe Tack tests are designed to measure the surface adhesion after a load pressure is applied to the probe. The details of these, and other standard adhesion tests are well described in the summary monograph by Abbott \cite{Abbott}. Much of the literature describes tests with a flat probe (ASTM D4541) \cite{Creton2020,Doi2018}, which is useful since the concepts of stress and strain are easily defined in such a geometry. However, we were unable to build a measuring device that would ensure the strict parallelity of the adhesive surface and the probe, and instead opted for the spherical probe: the `Probe Tack 2' method according to the Abbott classification. 

\begin{figure}
    \centering
   \includegraphics[width=0.4\textwidth]{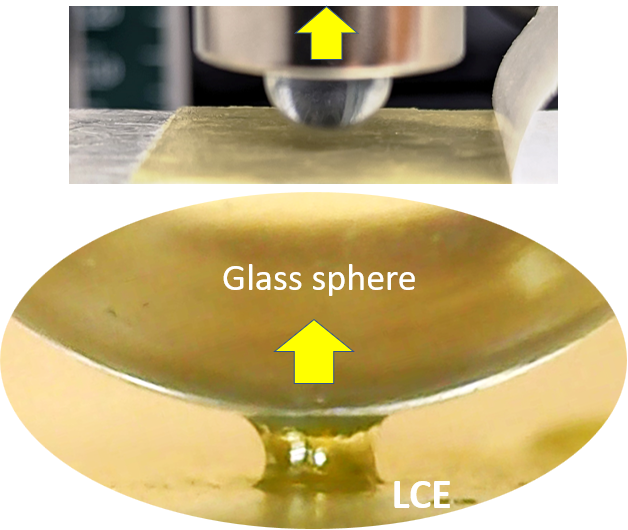}
    \caption{The illustration of the Probe Tack 2 test. Above: a photo of the glass sphere pressing into or pulled off the LCE adhesion tape facing up. Below: a zoomed-in region of contact during the pull-off.} 
   \label{fig2}
\end{figure} 

We used a spherical glass probe of 10\,mm diameter mounted on a vertical dynamometer frame   {(using the same Tinius Olsen ST1)}. The test protocol involved starting with a fresh LCE adhesive surface (the PET tape glued to a flat rigid base) and a cleaned glass sphere, which was lowered into the LCE layer until the fixed load of -0.5\,N was reached (this load level was determined empirically, by verifying that no damage was done to the LCE layer after the test). After that, the pre-determined dwell time (which we called `contact time' in the discussion)  was allowed, after which the probe was pulled up at a fixed rate of 10\,mm/min. It is well known that the adhesion strength depends on the pulling rate very strongly \cite{Abbott,Hiro2022}, but here we did not explore this test variable and kept it constant. The adhesion force is defined as the peak value of the tension force before the adhesive `neck' detaches from the probe, see Fig. \ref{fig2} for illustration, and the Supplementary Video 1 for detail.

\subsection*{90-degrees Peel test}

\begin{figure}
    \centering
   \includegraphics[width=0.4\textwidth]{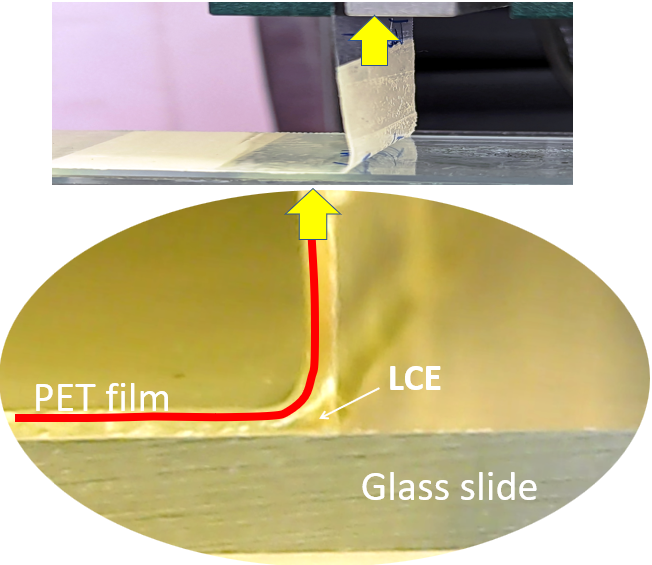}
    \caption{The illustration of the 90-degrees Peel test. Above: a photo of the glass slide with the PET-based LCE adhesive tape being pulled up at 90 degrees. Below: a zoomed-in region of contact during the pull-off.} 
   \label{fig3}
\end{figure}

The Peel Test followed the ASTM D3330 standard parameters and protocol, and we chose the 90-degrees peel geometry. Our `standard' 25\,mm-wide PET tape with LCE adhesive layer was pressed into the clean glass slide with a free end of the tape clamped into the vertical dynamometer frame. After a pre-determined contact time, the tape was pulled up   {(using the same Tinius-Olsen ST1 tensiometer)} with a rate of 100\,mm/min, recording the adhesion force. A crude method of horizontal movement of the flat glass slide base was devised to ensure the delaminating edge was always below the pulling clamp. See Fig. \ref{fig3} for illustration, and the Supplementary Video 2 for detail. 

\subsection*{Lap shear test}

The Lap Shear test followed the ASTM D3163 standard parameters. The PET tape was pressed into the clean glass slide with a free end of the tape clamped into the vertical dynamometer frame, while the free end of the glass clamped into the bottom clamp. In this geometry a shear deformation is imposed on the adhesive layer, with the shear strain proportional to the clamp displacement: $\varepsilon = x / d$, where the LCE layer thickness $d \approx$ 0.2\,mm.   {The same Tinius-Olsen ST1 tensiometer was used in this test.} 

\begin{figure}
    \centering
   \includegraphics[width=0.55\textwidth]{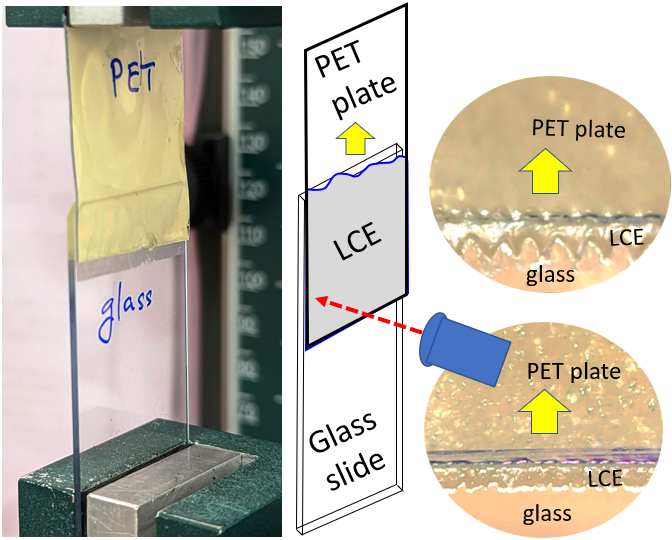}
    \caption{The illustration of the Lap Shear test. Left: a photo of the glass slide with the PET-based LCE adhesive tape being pulled up, inflicting the shear deformation in the LCE adhesive layer. Right: zoomed-in regions of the PET tape edge during the shear, illustrating different modes of LCE layer failure.} 
   \label{fig4}
\end{figure}

Here we had to make a variation on our PET tapes, because in this geometry we could not afford to have the PET backing stretch under the high forces involved -- and our `standard' PET of 23\,$\mu$m thickness was stretching easily. Therefore, for these tests we used PET `tape' of 0.2\,mm thickness, to act as backing for the same LCE layer, which was effectively rigid and did not provide any measurable deformation to contaminate the shear response of the LCE. 
Figure \ref{fig4} illustrates the test geometry, and the Supplementary Video 3 is helpful to visualize different modes of delamination in LCE layers of different crosslinking density.

%%%%%%%%%%%%%%%%%%%%%%%%%%%%%%%%%%%%%%%%%%%%%%%%%
\section{Results and Discussion}

\subsection{Probe Tack 2}
\begin{figure}[h!]
    \centering
   \includegraphics[width=1\textwidth]{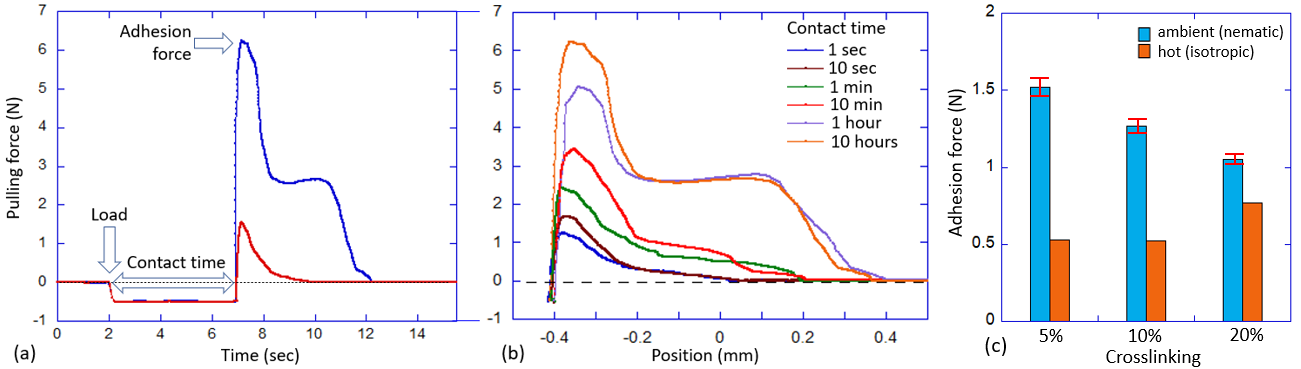}
    \caption{Probe Tack 2 results for a 10\,mm-glass sphere on LCE adhesive tape. (a) The summary illustration of a typical test output, showing a weak and a strong adhesion cases. (b) Comparing the pulling force response for the 5\%-crosslinked LCE layer after the increasing contact time. (c) The aggregate data for three different LCE layers, in the nematic and in the isotropic phase, at a fixed 5-second contact time. } 
   \label{fig5}
\end{figure}

The test summary in Fig. \ref{fig5}(a) illustrates the meaning of `adhesion force' in this context, and also of the `contact time', giving examples of two different adhesive responses: a weaker strength with direct delamination, and a higher strength with a neck forming on further pullout. We remind that at our chosen load of -0.5\,N there was no visible damage to the LCE layer after detachment. Figure \ref{fig5}(b), showing the data for the 5\%-crosslinked LCE, illustrates the strong dependence of adhesion on contact time: an almost 6-fold increase in the adhesion force over 10 hours of contact. 

We will discuss the time dependence in detail below, but first show the summary of this test in Fig. \ref{fig5}(c), for a specific 5-second value of contact time. It is clear that weaker-crosslinked LCE (with correspondingly longer network strands) has the higher adhesion force. The same test repeated at $T=80^\circ$C (in the isotropic phase of the LCE) shows a much lower adhesion. This thermal switching of PSA is fully reversible, with quite a small error also indicated in the plot. 

\begin{figure}
    \centering
   \includegraphics[width=0.85\textwidth]{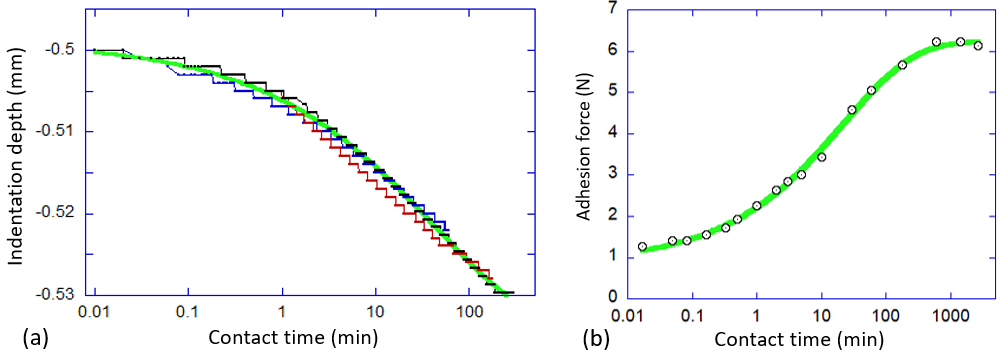}
    \caption{The effect of LCE relaxation in the Probe Tack 2 test; an example of 5\% crosslinked LCE. (a) Monitoring the indentation depth of the 10\,mm glass sphere held at a contact load of -0.5\,N. Three data sets are for the 1\,h, 3\,h and 10\,h contact time, with the solid green line presenting the fitted curve (see text). (b) The value of the ultimate (peak) adhesion force as a function of contact time, again with the fitted curve through the data.  } 
   \label{fig6}
\end{figure}

We see a strong dependence on contact time, and Fig. \ref{fig6} illustrates its different aspects in greater detail. First of all, we found that while the glass sphere is pressing into the polydomain LCE layer at constant force, the depth of indentation slowly increases. This is reminiscent of the studies of slow stress relaxation in polydomain LCE \cite{Clarke1998b}, and certainly relies on the same internal mechanism. The misaligned nematic domains that find themselves under stress tend to re-align along the local axis of principal extension. However, since neighboring domains are initially aligned differently, their individual shape change must be made mutually compatible. Adjusting this strongly correlated mechanical system is similar to the relaxation of the angle of repose in a sandpile, via decreasing local avalanches \cite{Nagel1989, Clarke1998b}, and it has been suggested to follow a logarithmic time dependence. We find it practically very difficult to distinguish the logarithmic time dependence from the stretched exponential with a low exponent, not without testing at very long times which our instruments did not permit. Figure \ref{fig6}(a) shows how the indentation depth increases in three separate experiments with different contact time, and the overlap of the curves is reassuring. We found the good fitting of this relaxation to follow the stretched-exponential equation
\begin{equation}
  d(t) = d_\mathrm{eq}+(d_0-d_\mathrm{eq})\cdot \left\{  1-\exp \left[ -\left( \frac{t}{\tau} \right)^{0.4} \right] \right\} \ , 
\end{equation}
where $d_0=0.5$\,mm is the initial indentation depth that we measure, the fitted saturation depth $d_\mathrm{eq} \approx 0.535$\,mm, and the fitted relaxation time constant is $\tau \approx$ 43.5\,minutes. 

Figure \ref{fig6}(b) shows a different aspect of this relaxation, plotting the set of peak values of adhesion force, as seen in Fig. \ref{fig5}(b), against contact time. Again, the logarithmic time axis allows visualization of both the short and the long time limits, the longest test we were able to carry out had a contact time of 45\,hours. This plot highlights the enormous, 6-fold increase in the adhesion strength, which cannot be explained merely accounting for the moderate increase in glass contact area that follows from the indentation increase. It is clear that the main contributing factor to this increase in adhesion strength is the slow re-adjustment of deformed nematic domains, which are able to achieve a much lower free energy state over time, and then `resist' having to lose this deep energy minimum on the sphere pullout. Importantly, a good fitting of this time dependence is achieved by the same stretched exponential:
\begin{equation}
  F(t) = F_\mathrm{eq}+(F_0-F_\mathrm{eq})\cdot \left\{  1-\exp \left[ -\left( \frac{t}{\tau} \right)^{0.4} \right] \right\} \ , 
\end{equation}
where $F_\mathrm{eq} = 6.23$\,N is the saturation force that we measure, the fitted initial adhesion strength (at zero contact time) is $F_0\approx 0.89$\,N, and the fitted relaxation time constant is now $\tau \approx$ 23\,minutes. We are not sure whether a small difference in the two time constants $\tau$ has any relevance, given so different types of data being fitted, but the fact that both relaxation laws follow the same stretched exponential law suggests a consistent mechanism of polydomain texture re-adjustment which is dominant in determining the strength of PSA.

\subsection{90-degree Peel}

The main difference in the peel test is that the adhesive tape is under no load during the contact time, unlike with the probe that is being pressed at constant load. The preparation of test samples is described in ASTM D3330 protocol: we use the standard 25\,mm wide adhesive tape on the glass slide, stick it using the prescribed 2\,kg roller, leave for a pre-determined contact time, and then peel at a high rate of 100\,mm/min. 

\begin{figure}[h!]
    \centering
   \includegraphics[width=\textwidth]{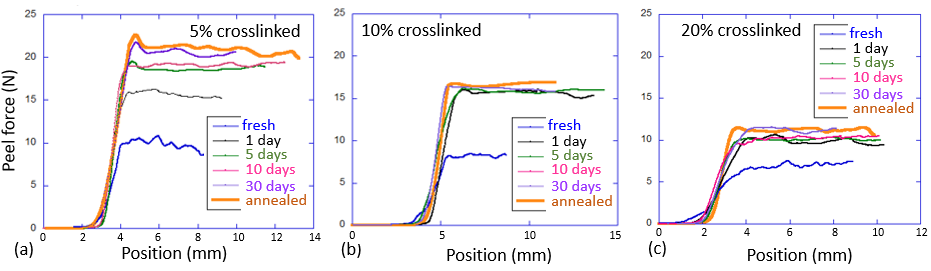}
    \caption{Peel Test results on a 25\,mm-wide adhesive tape, comparing the different LCE crosslinking densities: (a) 5\%, (b) 10\%, (c) 20\%. In each plot, different curves correspond to the different load-free contact time, as labelled in the graphs.  } 
   \label{fig7}
\end{figure}

Figure \ref{fig7} shows the essence of the test, with the x-axis representing the vertical position (which is equivalent to the time, at a constant pulling rate). After gathering some slack, the pulling force saturates at a constant level representing the adhesion force (per standard 25\,mm of the tape width).   {Note, that we follow the standard test notation of ASTM D3330, quoting the 'peel force' (in N) for the 25\,mm tape, but in many cases it may be instructive to look at the proper intensive parameter in units of N/m, or J/m${}^2$. In this way, the measurement of force =10\,N per 25\,mm in Fig.\ref{fig7} means 400\,N/m for the 20\%-crosslinked sample we label '1.2', or the force of 20\,N  means 800\,N/m for the 5\%-crosslinked sample '1.05'.}

As expected, and in agreement with the Probe Tack test (see the discussion \cite{Creton2020}), we find higher adhesion force for lower crosslinking densities, which in our case directly corresponds to the length of polymer strands in the network. For each crosslinking density, we again find a strong effect of the contact time. Interestingly, the saturation is reached much faster, within 24 hours for the highly crosslinked 20\% and 10\% LCE (samples 1.2 and 1.1), but takes over 5 days in the weakly crosslinked 5\% LCE sample 1.05. At the same time, the magnitude of the adhesion increase is just about 50\% for the 20\% LCE, while reaching over 120\% increase for the 5\% LCE adhesive film. Clearly, with longer network chains, the nematic polydomain structure is less constrained and is able to achieve a much `deeper' free energy state on re-adjustment near the surface -- but it takes a much longer time.

Another important point is also illustrated in Fig. \ref{fig7}. The ``annealed'' curve in all plots represents the test carried out on the freshly bonded tape, which was then annealed to the isotropic phase, and allowed to naturally cool back to the ambient temperature (all within 10 minutes after the first bonding). It is clear that such annealing enhances the adhesion strength to its maximum, which otherwise would be achieved after days of contact time at ambient temperature. We will return to the discussion of the underlying mechanism causing this effect later in the paper.

\begin{figure}
    \centering
   \includegraphics[width=0.85\textwidth]{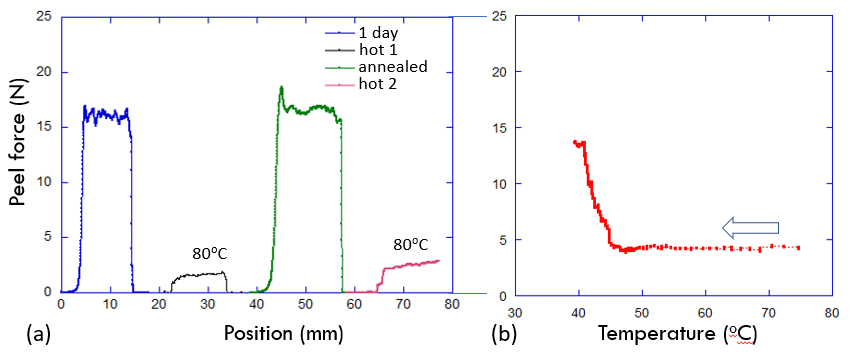}
    \caption{Phase effect on the LCE adhesion, using the example of 10\%-crosslinked LCE tape after 1-day bonding. (a) Repeated peel tests at ambient and at the high temperature show reproducibility of equilibrium adhesion strength in both phases. (b) The adhesion strength calibrated for continuously changing temperature on natural cooling of the tape during the peel test. } 
   \label{fig8}
\end{figure} 

Figure \ref{fig8} explores the switchable aspect of PSA in the LCE adhesive. Plot (a) combines two cycles of heating and cooling, using as an example the 10\% LCE tape. Starting at the ambient temperature (nematic phase of LCE) after 1-day contact time adhesion strength is reproducibly ca. 0.64\,N/mm, corresponding to the peel force of ca. 16\,N, in Fig. \ref{fig7}(b). When this tape is brought to a high temperature (isotropic phase), the continuing peel test shows the adhesion strength dropping to a reproducible ca. 0.08\,N/mm (peel force of 2\,N). When we then allow the sample to cool back to ambient temperature, the PSA tape condition becomes `annealed' (see the discussion around Fig. \ref{fig7}), and the adhesion strength returns to its equilibrium value. Heating again, returns to the weak adhesion in the isotropic phase. 
We could not repeat this test over many more cycles because the exposed LCE surface becomes contaminated after repeated adhesion-peel events. 

It was very difficult for us to test the accurate temperature dependence of adhesion strength, which would be desirable since we separately know the temperature dependence of the nematic order parameter and could make a more quantitative analysis. The best we could do is to heat the bonded tape, and calibrate its natural cooling rate (using the thermal camera to read the surface temperature, accepting the systematic error of the temperature gradient between the outer PET surface that we can view and the inner adhesive surface of LCE). Having calibrated the temperature vs. time in this way, we then started the peel test of the heated tape, allowing it to cool while being pulled. The result is presented in Fig. \ref{fig8}(b). Acknowledging all the experimental shortcomings, we nevertheless find a very clear reflection of the nematic phase transition, and a seemingly close correlation between the adhesion strength of the annealed LCE tape and the nematic order parameter that grows on cooling, before reaching the phase saturation.

\begin{figure}
    \centering
   \includegraphics[width=0.6\textwidth]{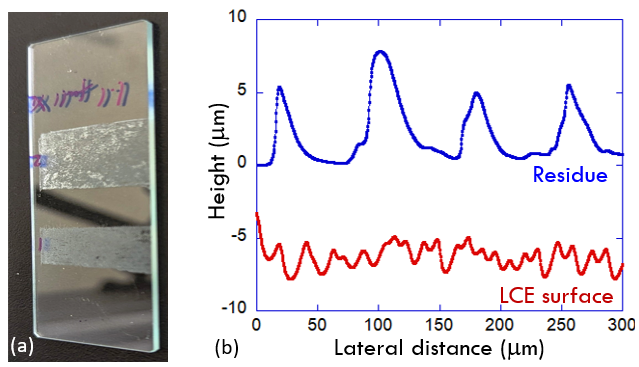}
    \caption{The residue left in high-temperature isotropic-phase peeling. (a) The photo of the glass slide left after the test of Fig. \ref{fig8}(a): as the tape was peeled off, first the residue-free debonding occurred in the ambient nematic phase, then the strip of residue was left on peeling at 80$^\circ$C, then again the clean debonding after the tape was cooled back to the ambient temperature, and then again the residue strip left after peeling at 80$^\circ$C. (b) The surface profilometer scans data, showing the characteristic dimensions of the natural free LCE surface, and of the residue layer left after isotropic peeling. } 
   \label{fig9}
\end{figure}

The peel tests at high temperature have revealed an unexpected phenomenon. We have to emphasize: in our lab, we have tested LCE adhesive tapes on many different surfaces (steel, ceramic tile, perspex plastic), and in all those cases we had no surface residue on peeling the PSA tape -- whether in the ambient nematic phase, or at high temperature in the isotropic phase. On the clean glass only, and at the high-temperature peeling only, but consistently across all our LCE materials, the tapes left a residue. Figure \ref{fig9}(a) illustrates this residue, and the lack of it when peeling the strongly bonded nematic LCE tape; in fact the photo is exactly of the sample presented in Fig. \ref{fig8}(a), with two segments of hot/isotropic peeling and two segments of ambient/nematic peeling. 

Figure \ref{fig9}(b) shows the surface topography of the natural LCE free surface, and of the residue layer discussed. The LCE surface profile is characteristic of the polydomain structure, where the regions aligned perpendicular to the surface elongate in the nematic phase and create the `hills'. However, it is obvious from comparison, that the residue left on debonding in the isotropic phase has nothing to do with these domains, as it has a very characteristic periodic length scale of ca. 80\,$\mu$m (much greater than the LCE domain size). In looking for an explanation for this effect, we have to return to Fig. \ref{fig1}(b) showing the intrinsic mechanical strength of our LCEs. It is clear that the isotropic phase of these networks is rather brittle, and breaking stress not exceeding 0.3\,MPa, in great contrast to the nematic phase. We have to assume that the (low) strength of isotropic adhesion on clean glass is still strong enough to cause the elastomer network to rupture instead of debonding (which is apparently not the case for other bonding surfaces we have tested). Given that we measured the isotropic adhesion strength (which we now realise is the cohesive failure) at 0.08\,N/mm and crudely taking the width of the stretched region to be 0.5\,mm (from the visual observation of peel boundary in Fig. \ref{fig3} and Supplementary Video 2), the resulting stress becomes ca. 0.16\,MPa. It is very easy to accept that a bit higher stress at the very front of the stretched zone reaches twice that value, which necessarily produces the cohesive crack, according to our tensile data. 

We will return to the discussion of residue, and of periodic structures developing at the debonding edge, in the next section on Lap Shear testing. 

\subsection{Lap Shear}

Lap shear test is a very different environment for the adhesive layer, and although it is an accepted alternative way of assessing the adhesion strength and features -- its results are not directly correlated with the Probe Tack or Peel tests \cite{Abbott}. For an LCE adhesive layer of 200\,$\mu$m thick, the shear strain of 500\% is reached by the 1\,mm movement of the top plate. Figure \ref{fig10}(a) gives the summary of lap shear results, and the comparison of three materials, in the nematic and isotropic phases. In the other tests, we saw an expected result that the weaker (5\%) crosslinked LCE layer has stronger adhesion. However, in shear geometry the response is different, and dominated by the cohesive failure of the bonded LCE layer. Both the 5\% and the 10\%-crosslinked LCE tapes reach the stress peak at a similar strain value of 700-800\%, similar to the tensile failure of these materials, see Fig. \ref{fig1}{(b). After this LCE network failure occurs (predictably, at a higher shear stress in the 10\% LCE), the response turns into the plastic flow of the top PET plate over the bottom glass plate. The fact that stress seems to diminish with further deformation (as opposed to the `constant stress' in the classical plastic creep) is simply due to the fact that less contact area is engaged. The `apparent' decrease of stress is an artefact of us dividing the diminishing force by the initial overlap area -- and when we take the proportionally diminishing area change into account, the proper plastic creep is obtained (at constant stress). The Supplementary Video 3 shows the detailed process of cohesive debonding viewed by the camera from above the overlap edge.  Note the continuous thin layer of residue left on the glass plate.

\begin{figure}
    \centering
   \includegraphics[width=0.85\textwidth]{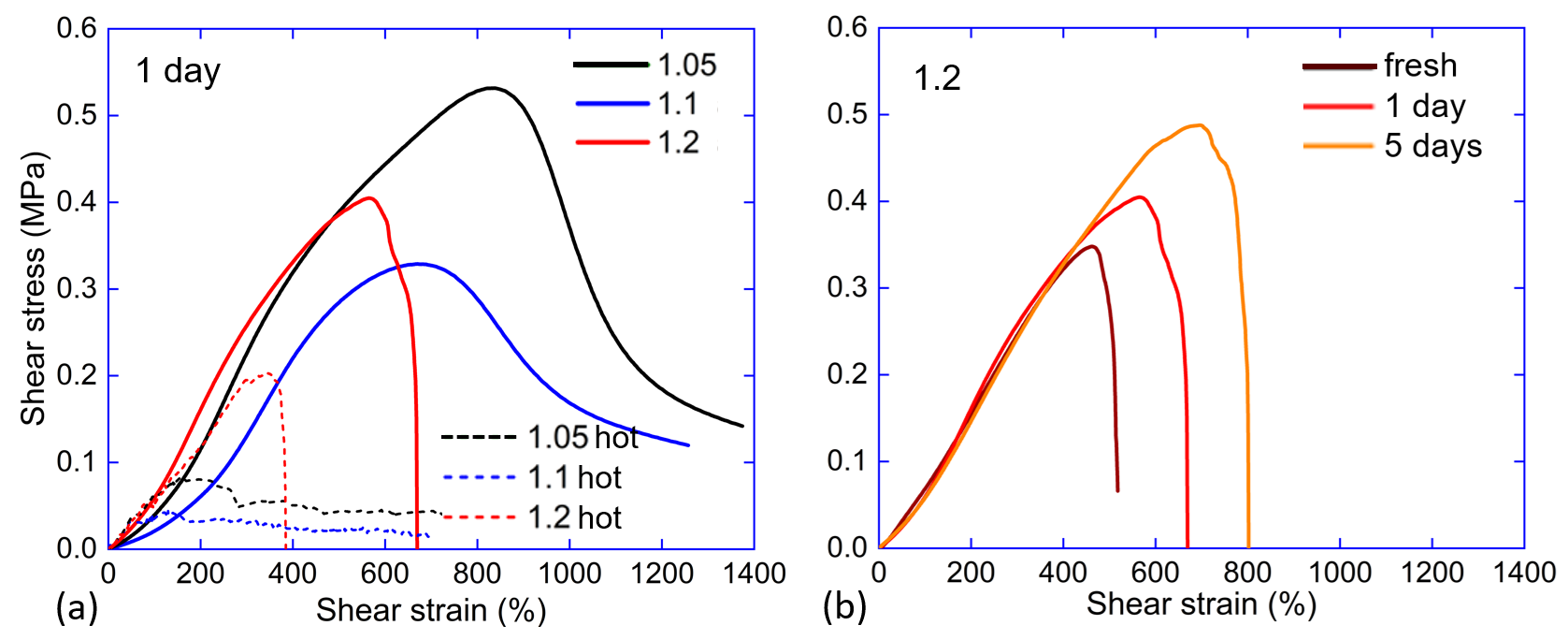}
    \caption{The Lap Shear testing of 25\,mm-wide tape with 10\,mm overlap. (a) Comparing the three crosslinking densities in shear test, in the nematic phase and in the high-temperature isotropic phase. There are two different modes of failure: the clean residue-free debonding for the 20\% LCE tape, and the cohesive failure resulting in the tape creep along the surface, and the residue layer, for the 10\% and 5\% LCE tapes. (b) A confirmation of the contact-time effect, for the 20\% LCE adhesive with its clean adhesive failure, but at different strength for different contact time.} 
   \label{fig10}
\end{figure}

In contrast, the strongly crosslinked 20\% LCE tape debonds by a clear adhesive failure without any residue, at a peak shear stress of 0.4\,MPa and strain ca. 500\%, evidently just before the LCE network failure. All LCE adhesive tapes debond at a low stress at high temperature, in the isotropic phase. But again, there is a contrast: the weaker crosslinked LCE tapes break across the LCE network, leaving the residue layer on glass -- while the stronger 20\% LCE continues to debond adhesively, only at a lower stress since we now know the adhesive strength is lower in the isotropic phase. Finally, we confirm the finding of other adhesion tests, that increasing the contact time on the glass surface makes the adhesion stronger, Fig. \ref{fig10}(b). This figure only presents the data for 20\% LCE tape as the only one actually testing the adhesion strength.  

It bears repeating the argument made earlier, about why the LCE network failure occurs in the thin layer, of approximately a domain size, above the flat adhesive surface with glass. Figure \ref{fig11} gives the sketch that illustrates this point. On pressured adhesion, the naturally rough free surface of LCE is forced to be flat, which distorts the first layer of nematic domains and forces them to start their slow nematic director re-adjustment to lower the elastic energy by exploring the local soft elastic trajectory for each domain. As was argued before, this is a slow process due to many mechanical constraints (barriers) that elastic domains must overcome to allow their director rotation, and is the cause of the adhesion strength increase over many days of contact. However, the local internal stresses that are left in the polydomain LCE add to the external shear stress imposed in the test, and the crack nucleation occurs in that thin layer near the contact plane. 

\begin{figure}
    \centering
   \includegraphics[width=0.45\textwidth]{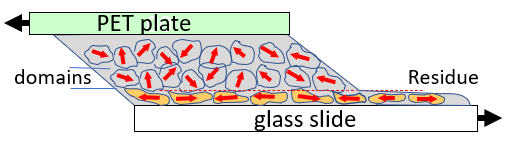}
    \caption{The sketch of the lap shear geometry of the polydomain LCE layer, highlighting the highly deformed zone of domains near the adhesive surface, which causes high internal stresses at the depth of domain size, and the resulting crack nucleating in that zone leading to the localized cohesive failure and the thin residue layer. } 
   \label{fig11}
\end{figure} 

\section{Conclusion}

In this paper, we carried out s series of industry-standard adhesion tests to reveal the mechanism of LCE adhesion, enhanced in the nematic phase and returning back to the ordinary elastomer levels in the isotropic phase. The essence of this mechanism is twofold, but in both cases based on the mobility of nematic director that is the core essence of LCEs \cite{LCEbook}. One factor causing a significant increase in the effective adhesive energy is the anomalous viscoelastic damping in nematic LCEs. This was alluded to in recent literature \cite{Hiro2022}, and is based on the theoretical link between the effective surface energy $\gamma$ and the dissipative loss (measured by $\tan \, \delta$) \cite{Persson2005}:
\begin{equation}
    \frac{\gamma}{\gamma_0} \approx \left( 1-\frac{2}{\pi} \int_0^{2\pi v/a} \frac{\sqrt{1-\omega a/2\pi v}}{\omega \tan \, \delta(\omega)} \mathrm{d}\omega \right)^{-1} , 
\end{equation}
where $\gamma_0$ is the `chemical' surface energy, $v$ is the speed of pulling (so that $v/a$ is the strain rate). The integral is over the frequency range determined by the strain rate. The reason why $\tan \, \delta$ is so high in the nematic phase of LCE is a separate matter, not very well understood but certainly related to the rotational viscosity of nematic director. However, the fact of increased viscoelastic dissipation in nematic LCE is not in question, and is one of the two reasons for the enhanced adhesion. This also suggests the route for material optimization: the higher the loss factor, the stronger the adhesion.

The other factor contributing to the high adhesion in the nematic phase is the significant reduction of the LCE elastic energy when a polydomain texture is deformed by pressing into the target surface, and the local director in each deformed domain is allowed to rotate, re-adjusting in correlation with its neighbors, exploring the soft elastic trajectory of deformation. The full re-adjustment of director takes a long time, following an unusual (possibly -- logarithmic) relaxation law due to the multiple correlation of local barriers in the random system. Once the low value of elastic energy is achieved, the bonded state becomes much preferred compared to the just-debonded state, where the condition for the energy minimum is removed. This suggests that polydomain LCE has a higher adhesion compared to an aligned monodomain LCE layer where no such soft elastic re-adjustment is possible.

Other promising routes for material optimization to further increase the adhesive strength, while retaining its reversible reduction in the isotropic phase (which could be achieved by heating above $T_\mathrm{NI}$ or by light stimulation if the LCE is made photo-sensitized), is by reduction of crosslinking density. However, this is not a simple matter, since an LCE network with less than ca. 5\% crosslinking becomes very weak and mechanically unstable. Therefore, the valid avenues are either by using the concept of `double network' to strengthen the LCE, or by forming a stronger network with dangling chain ends to provide the mobility and thus adhesion. Both of these routes are explored in the classical PSA systems \cite{hydrogel1,hydrogel2,dangling1,dangling2}, but we believe they have a great promise in the LCE field as well. 

%%%%%%%%%%%%%%%%%%%%%%%%%%%%%%%%%%%%%%%%%%%%%%%

\subsection*{Supplementary Material} 
Three videos are supplied as Supporting Information, available on request from the authors.
The following files are available.
\begin{itemize}
  \item Video S1: the probe tack experiment composition
  \item Video S2: the 90-deg peel test composition
  \item Video S3-5x: the lap shear test for the 5\% crosslinked LCE
  \item Video S3-20x: the lap shear test for the 20\% crosslinked LCE
\end{itemize}

\subsection*{Acknowledgements}
We are grateful to Professor Steven Abbott for many useful and encouraging conversations about PSA and adhesion testing. 
This work was supported by the European Research Council (H2020) AdG ``Active Polymers for Renewable Functional Actuators'' No. 786659, and PoC ``Reversible adhesion damping tapes of liquid crystalline elastomers'' No. 101100553). M.O.S. acknowledges support of the Royal Society. H.G. acknowledges support from the China Scholarship Council. 

%\section*{References}
%\bibliographystyle{naturemag}
%\bibliography{references}

\end{document}